\documentclass{emulateapj}
\slugcomment{Accepted in ApJ}

\begin{document}

\title{Modified Debye-H\"uckel Electron Shielding\\ and Penetration Factor}
\author{P. Quarati}
\affiliation{Dipartimento di Fisica - Politecnico di Torino, I-10129 Torino, Italy, and\\ Istituto Nazionale di Fisica Nucleare (INFN) - Sezione di Cagliari, I-09042 Monserrato, Italy.}
\email{piero.quarati@polito.it}
\author{A.M. Scarfone}
\affiliation{Dipartimento di Fisica - Politecnico di Torino, I-10129 Torino, Italy, and\\Istituto Nazionale di Fisica della Materia (CNR-INFM) - Unit\'a del Politecnico di Torino, I-10129 Torino, Italy.}\email{antonio.scarfone@polito.it}

\shorttitle{Modified Debye-H\"uckel Shielding}
\shortauthors{P. Quarati and A.M. Scarfone}

\begin{abstract}
Screened potential, modified by non standard electron cloud distributions responsible
for the shielding effect on fusion of reacting nuclei in astrophysical plasmas, is derived. The case of clouds
with depleted tails in space coordinates is discussed. The modified screened potential is obtained both from statistical mechanics arguments based on fluctuations of the inverse of the Debye-H\"uckel radius and from the solution of a Bernoulli equation used in generalized statistical mechanics. Plots and tables useful in evaluating penetration probability at any energy are provided.
\end {abstract}

\maketitle


\section{Introduction}

Nuclear fusion reaction cross sections and rates are sensitive to the
screening effect of the electron cloud around  reacting nuclei, an effect
that has been widely investigated both theoretically and experimentally
since the early works of Salpeter \citep{Salpeter1,Salpeter2}.\\
Different situations arise when fusion reactions take place: 1) in laboratory experiments, where a metal or gaseous target of a given element is bombarded by an ionic or charged particle beam, electrons are for the most part bound in atomic orbits and few of them can be considered free; 2) in stellar cores and other space and astrophysical plasmas where ions and nuclei are embedded in an electronic environment made by mainly free electrons; 3) in a deuterated metal and other solid-state  matrices where an impinging deuteron beam reacts with implanted deuterons.

--- In laboratory experiments, penetration through a screened Coulomb
potential at center of mass energy $E$ is shown to be equivalent to that
of bare nuclei at energy  $E+U_{\rm e}$ where $U_{\rm e}=Z_1\,Z_2\,e^2/
R_{\rm a}$ and $R_{\rm a}$ is the atomic radius or the radius of the
innermost electrons. $Z_1\,e$ and $Z_2\,e$ are the charges of the two reacting
nuclei and $U_{\rm e}$ is usually taken as constant in the evaluation of cross
sections and rates at any energy. Very often, at very low energy, fusion
cross sections measured in laboratory experiments are higher than the value
calculated by means of the usual Debye-H\"uckel (DH) screening factor. Stopping power of
the incoming beam and temperature reached after energy deposition are
important quantities for the correct measurement of the cross sections.
Modifications of electron distribution can be induced. The screening
effect must be evaluated to obtain the correct astrophysical factor at
very low energies \citep{Assenbaum,Carraro,Bracci,Shoppa,Strieder}.

--- In astrophysical plasmas, which that can be considered ideal, the free electrons
move around the reacting nuclei and occupy a sphere of
DH radius $R_{_{\rm DH}}=\sqrt{k\,T/(4\,\pi\,e^2\,n\,Z_\rho)}$
which is taken on the order of $R_{\rm a}$, with $n$ the particle density and
$Z_\rho=\sum_i(Z_i^2+Z_i)\,X_i/A_i$, where the sum is over all positive ions and
$X_i$ is the mass fraction of nuclei of type $i$. Only with a decreasing radius does the
screening effect become important. A screening factor of the rate can be derived
when the energy of the Gamow peak $E_{_{\rm G}}>U_{\rm e}$. It is given by the Debye
factor $f=\exp( U_{\rm e}/k\,T)$ with, this time, $U_{\rm e}=Z_1\,Z_2\,e^2/R_{_{\rm DH}}$ \citep{Rolfs,Ichimaru,Castellani,Opher}.
Recently, it has been clarified through numerous experimental observations
that the velocity distribution function of electrons (and possibly also of ions)
in stellar atmospheres, in space and astrophysical plasmas, may deviate from a Maxwell-Boltzmann distribution in the high-energy tail if non-local thermodynamic
effects are non-negligible \citep{Oxenius,Collins,Peyraund,Chevallier}.
In stellar atmospheres, atomic processes such as radiative and dielectronic
recombination show rates depending on deviations from Maxwellian
distributions of electrons \citep{Maero}. In stellar cores, signals of
possible deviations of ion distributions are evident and,
although small, should be considered because they are capable of sensibly influencing nuclear fusion reaction rates \citep{Ferro,Lissia}.

--- In deuterated metals or solid-state matter, the strong screening effect has still to be clearly understood and discussed, although a few interesting
descriptions have recently been brought forward to reproduce experimental results \citep{Raiola,Coraddu1,Coraddu2,Coraddu3,Kim}. The approach we are discussing here is very useful in understanding the fusion rates in this matter which could simulate some high-density astrophysical plasmas. However, this application deserves a separate detailed paper and we will not discuss this case here.

An important issue in the shielding of electrostatic potential in plasmas
concerns the investigation of non-linear charge screening effects that can
induce modifications in the DH potential, usually derived by
linearizing the Poisson equation \citep{Cravens,Gruzinov}.  One of the first studies which, on a microscopic basis, demonstrated deviations from DH factor can be found in \citep{Johnson}. [See also \citep{Shaviv1,Shaviv2,Shaviv3,Shaviv4,Chitanvis}]. The different approaches elaborated up to now are based on the assumption that electrons are distributed in space according to a Boltzmann factor. Few authors \citep{Bryant,Treuman,Leubner,Kim1,Rubab} have assumed a stationary $k$-Lorentz distribution to describe significant deviations from standard distributions, constructed from experimental distributions, due to the presence of enhanced high-energy tails. In a recent reference \citep{Rubab}, for instance, we can find quotations from many works where such distributions are reported. By analyzing energy profiles the authors derive an effective length smaller than the standard DH radius, depending on the $k$-parameter. This result is obtained by solving a Poisson equation in the linear weak approximation. The consequence must be an increased barrier penetration factor.

Our approach differs. We
assume that the electron cloud is spatially distributed following a generalized steady
state distribution of the $q$-type that reduces to an exponential distribution when the $q$ parameter (also known as the Tsallis entropic parameter) approaches the $q\to1$ limit \citep{Tsallis1,Tsallis2}. We refer the reader to Refs. \citep{Leubner1,Leubner2,Burlaga} for a detailed description of Tsallis generalized statistics and some of its applications to astrophysical problems. We then calculate the modified screening potential
by considering two different approaches. One concerns the use of a
generalized Poisson equation or Bernoulli equation as used by \citet{Tsallis2};
the other is based on super-statistics
\citep{Beck1,Beck2,Wilk2,Wilk1} considering fluctuations of an intensive
parameter (the inverse of the DH radius). This implies fluctuation in temperature
and density of the components of the plasma. Once we obtain the modified
potential, we  calculate penetration probability through that potential.
In this paper we limit ourselves to the value range $0<q<1$ (distribution in spatial coordinates with depleted tail cut at about $\langle1/R_{_{\rm DH}}\rangle$), leaving the case $q>1$ that describes a distribution in space coordinates with enhanced long-distance.

In the standard DH shielding approach, after linearization of the Poisson equation, the electrostatic screened potential behaves like $V_{_{\rm DH}}\sim r^{-1}\,\exp(-r/R_{_{\rm DH}})$ where $r$ is the coordinate with which to evaluate the DH potential. The assumptions made to derive the above relation are, among others things, that \citep{Cravens,Bellan}: plasmas are collisionless; the induced perturbation is slow and depends slowly on time (slowness); only electrostatic fields are present while induced fields are negligible; temperature is spatially uniform and the plasma remains in equilibrium during a perturbation; a temperature can always be defined; the number of particles inside the DH sphere is large and therefore fluctuations are small; although ions and electrons have a random thermal motion, perturbations induced around the equilibrium, responsible for small spatial variations of electrostatic potential, can be neglected.

Of course, if the real situation differs from the one imposed by one or more of
the assumptions reported above, the use of DH potential may induce errors
in the evaluation of the penetration factor and nuclear reaction rates.\\
Deviations from the conditions imposed by the assumptions are taken into account in
this work by choosing the inverse of DH length $1/R_{_{\rm DH}}$ as a
fluctuating parameter.\\
The electrostatic quantity $r\,V(r)$ is asymptotically given, in this case, by a power law instead of an exponential law because it must satisfy a differential equation, the Bernoulli equation (or a special case of it), that has power law functions as solutions.\\
We can derive the modified DH potential $V_ q(r)$ and, a posteriori,
the charge distribution $\rho_q$ as an asymptotically power law function. Also in the standard DH approach also two equations are needed, one coming from electromagnetism
(Poisson equation), the other from statistical mechanics (Boltzmann factor).\\
Penetration factor $\Gamma(E)$ can be calculated by means of the WKB
approach using the modified DH potential. We obtain results that differ from
the ones calculated with standard DH potential and that will be useful in the
interpretation of experimental results on atomic and nuclear rates in
several astrophysical systems and processes. We also derive equivalent
energy $U_q$ that we give by an interpolating analytical expression, a plot
and a tabulation. Energy $U_q$ results a function of the variable
$D/E$, where $D$ is defined by $D=Z_1\,Z_2\,e^2\,\langle1/R_{_{\rm DH}}\rangle$.
It is easy to observe that $U_q$ depends, for a wide range of values of $D/E$,
on $1/\sqrt{k\,T}$ and $\sqrt{n}$.

In Section 2, we explain how to derive the modified potential on the basis of
super-statistics arguments with the inverse DH length as a fluctuating
parameter; we then introduce a nonlinear differential equation, to be
associated with the Poisson equation, whose solution coincides with the potential
derived directly from the super-statistics approach.\\
In Section 3, we derive the penetration factor that can be used for evaluation
of nuclear fusion rates in astrophysical plasmas and indicate the range of validity of approximations
adopted. In Section 4, we discuss some representative examples whilst, in Section 5, we report our conclusions.

\section{Modified Debye-H\"uckel potential}

Combining the Gauss law and the relation that links the electrostatic field to
the electric potential $V(r)$ of a point test unitary charge at the origin
in a vacuum, from the Poisson equation one obtains the pure Coulomb potential.
After a sufficiently long time, electrons and ions rearrange themselves as a
response to the forces on them. Ion density eventually remains uniform while
electron density near the test charge increases. At the new thermal equilibrium,
the distribution of electrons in an electrostatic field is assumed to be given
by the well-known Boltzmann factor. Assuming the Boltzmann factor for all the
particles, after linearization and using neutrality condition, the Poisson
equation can be written as
\begin{equation}
{1\over r}\,{d^2\over d r^2}\Big(r\,V(r)\Big)={1\over R_{_{\rm DH}}^2}\,V(r) \ ,\label{poi}
\end{equation}
with solution given by the DH potential and the charge density
$\rho_{_{\rm DH}}$ expressed as
\begin{equation}
\rho_{_{\rm DH}}\sim-{1\over r\,R_{_{\rm DH}}}\,\exp\left(-{r\over R_{_{\rm DH}}}\right) \ .
\end{equation}
When one or more of the linearity constraints are violated or relaxed, a different description of the screening is required.

If we assume that nonlinear effects produce fluctuations on the inverse DH
radius, by following the approach usually developed by super-statistics for inverse
temperature $\beta=1/k\,T$ \citep{Beck1,Beck2}, we can describe the
plasma around the test charge as made of cells where  $R_{_{\rm DH}}$ is
approximately constant and the system can be described by ordinary statistical
mechanics, in this case by the exponential (Boltzmann) factor $\exp(-r/R_{_{\rm DH}})$.
In the long term run the system is described by a spatial average over the mean of
fluctuating quantity $1/R_{_{\rm DH}}$. Fluctuation of the inverse DH radius also means fluctuation of the plasma
parameter given by
\begin{equation}
\gamma={1\over n\,R_{_{\rm DH}}^3}=\left(4\,\pi\,e^2\,Z_\rho\over k\,T
\right)^{3/2}\,\sqrt{n} \ .
\end{equation}

With few changes, we follow the approach by Wilk and W{\l}odarczyk for the case  of distributions with depleted tails ($q<1$) \citep{Wilk1,Wilk2}. Here we focus our attention on $q<1$ distribution because this shows a depleted tail with a cut-off, that is, the spatial distribution we assume for the electrons.\\
We assume that a certain variable $r$ of the system is limited between $0$ and
$[(1-q)\,\lambda_0]^{-1}$ where $\lambda_0$ is a constant parameter.\\
We define the function
\begin{equation}
{\cal F}_{q<1}(r,\,\lambda_0)=C_q\,\int\limits_0\limits^\infty f_{q<1}
\left(r,\,\lambda;\,\lambda_0\right)\,\exp\left(-\lambda\,r\right)
\,d\lambda \ ,
\end{equation}
where $C_q$ is a normalization factor and $f_{q<1}(r,\,\lambda;\,\lambda_0)$ is the probability density to observe a certain
value $\lambda$ of the system which is spread around the value $\lambda_0$. The
expression we choose for $f_{q<1}(r,\,\lambda;\,\lambda_0)$ is a gamma distribution
\begin{equation}
f_{q<1}(r,\,\lambda;\,\lambda_0)={A_q(r;\,\lambda_0)^{1\over1-q}\over
\Gamma\left({1\over1-q}\right)}\,\lambda
^{{1\over1-q}-1}\,\exp\Big(-{\lambda\,A_q(r;\,\lambda_0)}\Big) \ ,
\end{equation}
where
\begin{equation}
A_q(r;\,\lambda_0)={2-q\over 1-q}\,\lambda_0^{-1}-r \ ,
\end{equation}
and $q$ is the entropic Tsallis parameter.\\
Inserting the function  $f_{q<1}(r,\,\lambda;\,\lambda_0)$ into ${\cal F}_{q<1}(r;\,\lambda_0)$ we
obtain the normalized power law distribution
\begin{equation}
{\cal F}_{q<1}(r;\,\lambda_0)=\lambda_0\,
\left(1-{1-q\over2-q}\,\lambda_0\,r\right)^{1\over1-q} \ .
\end{equation}
Average value and variance of $\lambda$ depend on variable $r$ according to
\begin{equation}
\overline\lambda={1\over(1-q)\,A_q(r;\,\lambda_0)} \ ,\hspace{10mm}
\overline{\lambda^2}={2-q\over\big[(1-q)\,A_q(r;\,\lambda_0)\big]^2} \ ,
\end{equation}
where $\overline {x(r)}=\int x(r,\,\lambda)\,f_{q<1}(r,\,\lambda;\,\lambda_0)\,d\lambda$ is evaluated by means of distribution
$f_{q<1}(r,\,\lambda;\,\lambda_0)$.\\
However, the
relative variance depends only on $q$
\begin{equation}
\omega={\overline{(\lambda^2)}-(\overline\lambda)^2\over
(\overline\lambda)^2}=1-q \ .
\end{equation}
We remark that quantity $\lambda_0$ coincides with the spatial average
of $\overline\lambda$, that is
\begin{equation}
\langle\lambda\rangle=\lambda_0 \ ,
\end{equation}
where $\langle x\rangle=\int\overline{x(r)}\,{\cal F}_{q<1}(r;\,\lambda_0)\,dr$ is evaluated by means of distribution
${\cal F}_{q<1}(r;\,\lambda_0)$ which is the weighted average of the exponential
(or Boltzmann-like) factor $\exp(-\lambda\,r) $ with weight equal to
$f_{q<1}(r,\,\lambda;\,\lambda_0)$ and coincides with the Laplace transform of $f_{q<1}(r,\,\lambda;\,\lambda_0)$.\\
Some special limiting cases are
\begin{equation}
f_{q\to1}(r,\,\lambda;\,\lambda_0)=\delta(\lambda-\lambda_0) \ .
\end{equation}
and
\begin{equation}
{\cal F}_{q\to1}(r;\,\lambda_0)=\lambda_0\,\exp(-\lambda_0\,r) \ ,
\end{equation}
the ordinary Boltzmann-like factor, with $\overline{(\lambda^2)}=(\overline\lambda)^2$ and
\begin{equation}
f_{q\to0}(r,\,\lambda;\,\lambda_0)=A_0(r;\,\lambda_0)\,
\exp\Big(-\lambda\,A_0(r;\,\lambda_0)\Big) \ ,
\end{equation}
where $A_0(r;\,\lambda_0)=2/\lambda_0-r$ and
\begin{equation}
{\cal F}_{q\to0}(r;\,\lambda_0)=\lambda_0\,(1-\lambda_0\,r) \ ,
\end{equation}
a linear function, with
$\overline{(\lambda^2)}=2\,(\overline\lambda)^2$.\\
Let us now identify
functional ${\cal F}_{q<1}(r;\,\lambda_0)$ with quantity $r\,V_q(r)$ and substitute $\lambda$ with $1/R_{_{\rm DH}}$, so that $\lambda_0=\langle1/R_{_{\rm DH}}\rangle$ coincides with the spatial average of the DH radius fluctuation. By posing
\begin{equation}
\zeta_q=(2-q)\,\langle1/R_{_{\rm DH}}\rangle^{-1} \ ,
\end{equation}
a characteristic length of the system under inspection which reduces to $R_{_{\rm DH}}$ in the $q\to1$ limit, we
obtain the following expression
\begin{equation}
V_q(r)={1\over r}\,\left(1-(1-q)\,{r\over\zeta_q}\right)^{1\over1-q} \ ,\label{qpot}
\end{equation}
which, for $q\to1$ reduces to the standard DH potential. We also have
charge distribution
\begin{equation}
\rho_q(r)\sim-{1\over (2-q)\,r\,\zeta_q^2}\,\left(1-(1-q)\,{r\over\zeta_q}\right)^{1\over1-q} \ ,\label{qcharge}
\end{equation}
which for $q\to1$ reduces to $\rho_{_{\rm DH}}$, the charge
distribution of the DH approximation.\\Therefore, by considering $1/R_{_{\rm DH}}$ subject to fluctuations described by a gamma distribution, quantity $r\,V(r)$ related to the potential energy barrier is modified from the exponential DH expression to a power-like law,
typical of generalized $q<1$ distribution. When we can establish the functional
relation between density $n$ and temperature $k\,T$, as, for instance in a solar-like star, where quantity $n/(k\,T)^3$ is constant along the star profile \citep{Ricci}, by means of the relative variance we can establish a link between inverse DH radius and temperature fluctuations and parameter $q$ through
\begin{equation}
{\Delta(1/R_{_{\rm DH}})\over1/R_{_{\rm DH}}}=
{\Delta\sqrt{n/k\,T}\over \sqrt{n/k\,T}}={\Delta(k\,T)\over
k\,T}=(1-q)^{1/2} \ .
\end{equation}
Diffusion of matter between layers with different temperatures induces local temperature
fluctuations and density perturbations; fluctuations around an equilibrium or a steady-state matter profile induce fluctuations of quantity $1/R_{_{\rm DH}}$ in the DH sphere, particularly in the regions where the number of particles inside the DH sphere is small. Density and temperature fluctuations do not alter the macroscopic plasma parameters and must agree with the requirements of the constraints imposed by the macroscopic observations. In the solar interior, \citet{Gruzinov} have evaluated that the DH radius is about $2\cdot10^{-9}cm$, therefore containing a small number of particles with a corresponding
non-negligible particle fluctuation.

Let us now justify expressions (\ref{qpot}) and (\ref{qcharge}) by
means of another approach that considers a generalized version of the Poisson equation. In fact, in the standard case, the DH potential can be obtained from the solution of the
second order differential equation of
the type
\begin{equation}
{dy\over dr}=a\,y \ , \hspace{10mm} {\rm or} \hspace{10mm} {d^2y\over dr^2}=a^2\,y \ ,
\end{equation}
with $y\equiv r\,V(r)=\exp(a\,r)$.\\ We replace the above linear equation with the following one
\begin{equation}
{dy\over dr}=a_q\,y^q \ , \hspace{10mm} {\rm or} \hspace{10mm} {d^2y\over dr^2}=q\,a_q^2\,y^{2\,q-1} \ ,
\end{equation}
with
\begin{equation}
y=\exp_q(a_q\,r)\equiv\Big[1-(1-q)\,a_q\,r\Big]^{1\over1-q} \ ,
\end{equation}
where $q$ is a real number parameter and coincides with the Tsallis parameter. For $q\to1$
we obtain $\exp_1(a\,r)\equiv \exp(a\,r)$.\\
To be explicit, for our case we generalize Eq. (\ref{poi}) into
\begin{equation}
{d^2\over dr^2}\Big(r\,V_q(r)\Big)={q\over\zeta_q^2}\,\Big(r\,V_q(r)\Big)^{2\,q-1} \ ,
\end{equation}
which is a generalized Poisson equation whose solution coincides with Eq. (\ref{qpot}).\\ We report, for completeness, that by considering the Bernoulli equation
introduced by \citet{Tsallis2}
\begin{equation}
{dy\over dr}=a_1\,y+a_q\,y^q \ ,
\end{equation}
or
\begin{equation}
{d^2y\over dr^2}=a_1^2\,y
+(1+q)\,a_1\,a_q\,y^q+q\,a_q^2\,y^{2\,q-1} \ ,
\end{equation}
the solution is
\begin{equation}
y=\Bigg[e^{(1-q)\,a_1\,r}+{a_q\over a_1}\,\Big(e^{(1-q)\,a_1\,r}-1\Big)\Bigg]^{1\over1-q} \ .
\end{equation}
By posing $y=r\,V(r)$ we obtain
$$r\,V(r)=\Big[1+(1-q)\,a_q\,r\Big]^{1\over1-q}$$
when $a_1=0$ and $r\,V(r)=\exp(a_1\,r)$ when $a_q=0$.

\section{Penetration probability}

We calculate penetration probability through the repulsive barrier, one of the terms we need for evaluation of fusion reaction rates, using the WKB approach and following  \citet{Bahcall}.\\
The fusion cross section of two isolated reacting nuclei is written as
\begin{equation}
\sigma(E)={S(E)\over E}\,e^{-2\,\pi\,\eta(E)} \ ,
\end{equation}
where $S(E)$ is the astrophysical factor, $E$ the center of mass energy of the fusing
nuclei of charge $Z_1\,e$ and $Z_2\,e$ colliding with relative velocity
$v/c=\sqrt{2\,E/(\mu\,c^2)}$, reduced mass $\mu$ and
\begin{equation}
\eta(E)={1\over\hbar\,c}\,{Z_1\,Z_2\,e^2\over \sqrt{E}}\,{\sqrt{{1\over2}\,\mu\,c^2}} \ .
\end{equation}
First of all we define the  penetration factor when  the pure Coulomb electrostatic potential energy barrier $\widehat V_{_{\rm C}}(r)=Z_1\,Z_2\,e^2\,V_{_{\rm C}}(r)$ occurs, that is, when the reacting nuclei are isolated:
\begin{eqnarray}
\nonumber
\Gamma_{_{\rm C}}(E)&=&e^{-2\,\pi\,\eta(E)}\\
\nonumber
&=&\exp\left\{-{2\over\hbar\,c}\int\limits_0
\limits^{r_{\rm c}}\left[2\,\mu\,c^2\,\left(\widehat V_{_{\rm C}}(r)-E\right)\right]^{1/2}\,dr\right\} \ ,\\
&&\label{gc}
\end{eqnarray}
where $r_{_{\rm C}}$ is the classical turning point whose value is fixed by the relation
$\widehat V_{_{\rm C}}(r_{_{\rm C}})=E$.

Secondly, by using the standard DH potential and still taking the turning point
$r_{_{\rm DH}}=r_{_{\rm C}}$, a relation valid only for $E>D$, for small values of
$r_{_{\rm C}}/R_{_{\rm DH}}$ we have
\begin{equation}
\Gamma_{_{\rm DH}}(E)=\exp\left[-\pi\,\left(2+{r_{_{\rm C}}\over R_{_{\rm
DH}}}\right)\,\eta(E)\right] \ .\label{gdh}
\end{equation}
If we consider the rates instead of the cross sections, factor
$\exp(-\pi\,\eta\,r_{_{\rm C}}/R_{_{\rm
DH}})$ can be evaluated at most probable energy $E_0$ in such a way that the rate can be factorized as the product of $\Gamma_{_{\rm C}}(E)$ times a factor fixed at $E=E_0$.

Finally, we consider the deformed DH potential energy barrier
\begin{equation}
\widehat V_q(r)={D\over r\,\big\langle{1/R_{_{\rm
DH}}}\big\rangle}\,\left[1-(1-q)\,{r\over\zeta_q}\right]^{1\over1-q} \ ,
\end{equation}
and the penetration factor
\begin{equation}
\Gamma_q(E)=\exp\left\{-{2\over\hbar\,c}\int\limits_0\limits^{r_q}
\left[2\,\mu\,c^2\,\left(\widehat V_q(r)-E\right)\right]^{1/2}\,dr\right\} \ ,\label{gq}
\end{equation}
where $r_q$ must be derived from relation $\widehat V_q(r_q)=E$.\\
Because we consider $q<1$ potential energy $\widehat V_q(r)$ has a cutoff and, as a
consequence, $0<r<\zeta_q/(1-q)$.
We write Eq. (\ref{gq}) as
\begin{equation}
\Gamma_q(E)=\exp\Big(-2\,\pi\,\eta(E)\,\tau_q\Big) \ ,
\end{equation}
where function $\tau_q$, which depends on quantity $D/E$, goes to one for $D/E\to0$.\\The evaluation of $r_q$ and $\Gamma_q(E)$ can be worked out only numerically although, for small deformation $(q\approx1)$, the penetration factor can be worked out analytically and is given as a product of $\Gamma_{_{\rm C}}(E)$ times a correction factor. However we do not report here the expression for simplicity's sake. In the case $q=0$, which represents the greatest deformation with respect to the exponential function, $\tau_q$ has the simplest analytical solution
\begin{equation}
\tau_{q=0}={E\over E+D} \ .
\end{equation}
\begin{figure}	
\hspace{-10mm}\includegraphics[width=110mm]{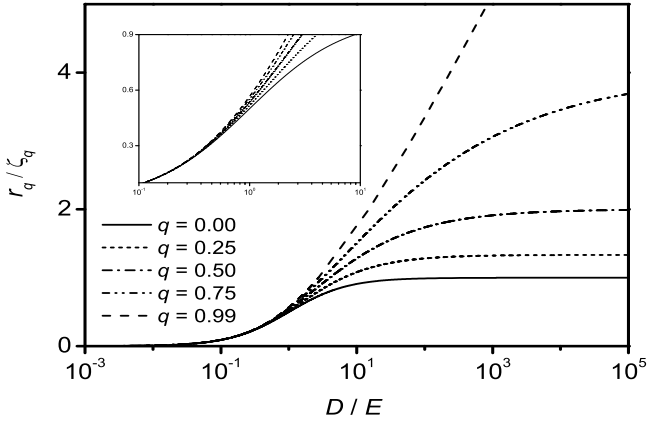}
\caption{Log-linear plot of the quantity $r_q/\zeta_q$ as a function of $D/E$ for several values of $q$ (in the insert the
region $10^{-1}<D/E<10$ is expanded).}
 \end{figure}
In Figure 1, we report the plot of quantity $r_q/\zeta_q$ as a function of $D/E$ for a few values of $q$ between zero and one.\\
In Figure 2, quantity $1-\tau_q$ is plotted as a function of $D/E$.\\
Penetration of potential energy barrier $\widehat V_q(r)$ at energy $E$ is equivalent to penetration of the pure Coulomb barrier at an effective energy $E+U_q$ where
\begin{equation}
U_q=\left({D\over r_q\,\big\langle{1/R_{_{\rm
DH}}}\big\rangle}-E\right) \ ,
\end{equation}
is a function of $D/E$.\\
\begin{figure}	
\hspace{-10mm}\includegraphics[width=110mm]{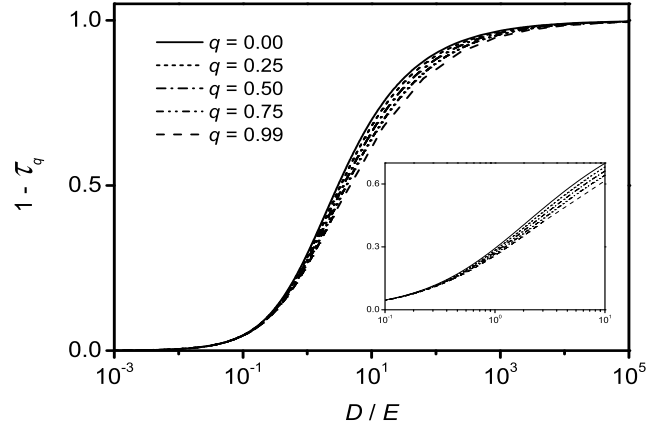}
 \caption{Log-linear plot of the quantity $1-\tau_q$ as a function of $D/E$ for several values of $q$ (in the insert the region $10^{-1}<D/E<10$ is expanded).}
 \end{figure}
We have also calculated numerically equivalent energy $U_q$. The behavior  of $U_q/D$, as a function of $D/E$, for a few values of $q$ is plotted in Figure 3. Finally, a quantitative comparison of quantities $r_q/\zeta_q$, $U_q/D$ and $1-\tau_q$, corresponding to the values of $q$ depicted in the figures can be obtained from the numerical Table 1.\\
\begin{figure}	
\hspace{-10mm}\includegraphics[width=110mm]{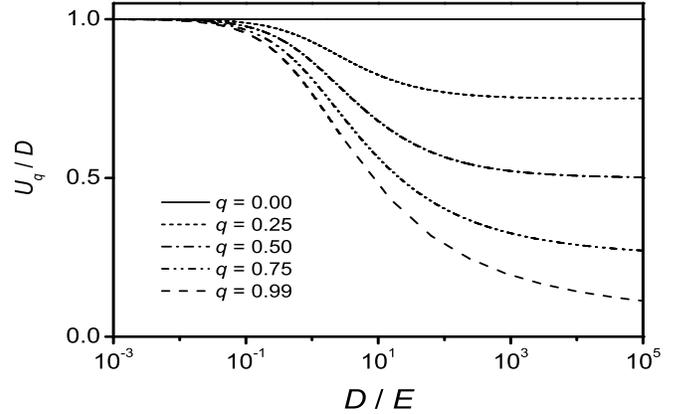}
 \caption{Log-linear plot, in arbitrary unities, of $U_q/D$ as a function of $D/E$ for several values of $q$.}
 \end{figure}
For $q=0$ potential $V_{q=0}$ is cut at $r=2\,\langle1/R_{_{\rm
DH}}\rangle^{-1}$ and we have $U_q=D$ for any value of energy $E$. In the other cases of $0<q<1$ we have that: at high energy $E>D$, $U_q$ approaches $D$. At low energy, $E<D$, $U_q$ approaches the value $(1-q)\,D$, meaning that $U_1$ (DH exponential potential) approaches zero. Therefore, when the electron distribution is a deformed or generalized distribution of the $q$-type, with a cut-off at $r=(2-q)/[(1-q)\,\langle1/R_{_{\rm
DH}}\rangle]$, the penetration factor is enhanced even at very low energy, except for the case $q=1$. Enhancement depends on $U_q$ which is an energy proportional to $1/\sqrt{k\,T}$ and $\sqrt{n}$.\\
\begin{deluxetable*}{lccc||ccc||ccc||ccc}
\tabletypesize{\scriptsize}
\tablecaption{Few numerical values of the quantities $r_q/\zeta_q$, $U_q/D$ and $1-\tau_q$, in the energy range $0.1\leq D/E\leq30$ for some values of $q$.}
\tablewidth{173mm}
\tablehead{\colhead{$D/ E$} & \colhead{$r_q/\zeta_q$} & \colhead{$U_q/ D$} & \colhead{$1-\tau_q$} & \colhead{$r_q/ \zeta_q$} & \colhead{$U_q/ D$} & \colhead{$1-\tau_q$} & \colhead{$r_q/ \zeta_q$} & \colhead{$U_q/ D$} & \colhead{$1-\tau_q$} & \colhead{$r_q/ \zeta_q$} & \colhead{$U_q/ D$} & \colhead{$1-\tau_q$}\\
}
\startdata
 & & $q=0.25$ & & & $q=0.50$ & & & $q=0.75$ & & & $q=0.099$ & \\
0.10 & 0.091 & 0.988 & 0.046 & 0.091 & 0.977 & 0.045 & 0.091 & 0.966 & 0.045 & 0.091 & 0.956 & 0.045\\
0.20 & 0.167 & 0.978 & 0.085 & 0.167 & 0.958 & 0.084 & 0.168 & 0.938 & 0.083 & 0.168 & 0.920 & 0.082\\
0.30 & 0.232 & 0.969 & 0.120 & 0.233 & 0.941 & 0.118 & 0.235 & 0.915 & 0.116 & 0.236 & 0.891 & 0.114\\
0.40 & 0.288 & 0.961 & 0.151 & 0.291 & 0.927 & 0.148 & 0.294 & 0.894 & 0.145 & 0.297 & 0.866 & 0.142\\
0.50 & 0.338 & 0.955 & 0.178 & 0.343 & 0.914 & 0.174 & 0.347 & 0.877 & 0.170 & 0.351 & 0.844 & 0.166\\
0.60 & 0.382 & 0.948 & 0.203 & 0.389 & 0.902 & 0.198 & 0.395 & 0.861 & 0.193 & 0.401 & 0.825 & 0.188\\
0.70 & 0.421 & 0.942 & 0.226 & 0.430 & 0.892 & 0.220 & 0.439 & 0.846 & 0.214 & 0.447 & 0.807 & 0.208\\
0.80 & 0.457 & 0.937 & 0.246 & 0.468 & 0.882 & 0.239 & 0.479 & 0.834 & 0.232 & 0.489 & 0.792 & 0.226\\
0.90 & 0.489 & 0.932 & 0.265 & 0.503 & 0.874 & 0.257 & 0.517 & 0.822 & 0.250 & 0.529 & 0.778 & 0.243\\
1.00 & 0.518 & 0.928 & 0.283 & 0.535 & 0.866 & 0.274 & 0.552 & 0.811 & 0.266 & 0.566 & 0.765 & 0.258\\
2.00 & 0.716 & 0.896 & 0.407 & 0.763 & 0.809 & 0.393 & 0.809 & 0.735 & 0.380 & 0.850 & 0.675 & 0.367\\
3.00 & 0.826 & 0.876 & 0.482 & 0.902 & 0.774 & 0.465 & 0.977 & 0.689 & 0.449 & 1.047 & 0.621 & 0.434\\
4.00 & 0.898 & 0.863 & 0.534 & 1.000 & 0.750 & 0.516 & 1.102 & 0.657 & 0.498\ & 1.198 & 0.584 & 0.482\\
5.00 & 0.949 & 0.852 & 0.572 & 1.073 & 0.731 & 0.554 & 1.200 & 0.633 & 0.535 & 1.321 & 0.556 & 0.518\\
6.00 & 0.988 & 0.844 & 0.603 & 1.131 & 0.717 & 0.583 & 1.281 & 0.613 & 0.564 & 1.426 & 0.534 & 0.546\\
7.00 & 1.019 & 0.838 & 0.627 & 1.179 & 0.705 & 0.608 & 1.349 & 0.598 & 0.588 & 1.517 & 0.516 & 0.570\\
8.00 & 1.043 & 0.832 & 0.648 & 1.219 & 0.695 & 0.628 & 1.408 & 0.584 & 0.608 & 1.597 & 0.500 & 0.590\\
9.00 & 1.064 & 0.828 & 0.665 & 1.253 & 0.686 & 0.646 & 1.461 & 0.573 & 0.626 & 1.670 & 0.487 & 0.607\\
10.00 & 1.081 & 0.824 & 0.680 & 1.283 & 0.679 & 0.661 & 1.507 & 0.563 & 0.641 & 1.735 & 0.476 & 0.622\\
15.00 & 1.140 & 0.810 & 0.733 & 1.390 & 0.652 & 0.715 & 1.684 & 0.526 & 0.695 & 1.996 & 0.434 & 0.676\\
20.00 & 1.174 & 0.801 & 0.766 & 1.459 & 0.635 & 0.749 & 1.806 & 0.503 & 0.730 & 2.188 & 0.406 & 0.711\\
\enddata
\end{deluxetable*}
We have also interpolated the function $U_q/D$ (for values $0<q<0.8$) which is well described by the following analytic function
\begin{equation}
{U_q\over D}=1-q+q\,\left[1-(1-q)\,a\,\left(D/E\right)^b\right]^{-{c\over1-q}} \ ,\label{interpo}
\end{equation}
where parameters $a(q)$, $b(q)$ and $c(q)$ are given by
\begin{eqnarray}
\nonumber
&&a(q)=0.484012+1.984961\,q+42.933001\,q^8 \ ,\\
\nonumber
&&b(q)=0.958079-0.129717\,q-0.038993\,q^5+\\
&&\hspace{10.5mm}0.804107\,q^6 \ ,\\
\nonumber
&&c(q)=0.760938-1.272798\,q+0.493549\,q^2 \ .
\end{eqnarray}
The interpolating function allows us to find the value of $U_q(E)$ once we know the temperature and electron density, fixed at $Z_1$, $Z_2$, and energy $E$ for a certain value of $q$.
In the range $0<D/E\leq10^2$, for $0.8<q<1$, function (\ref{interpo}) still gives a good interpolation of $U_q/D$ but with more complicated relationships of $a(q)$, $b(q)$ and $c(q)$. We omit the details.\\
Calculation of nuclear fusion rates (quantities weighted over the reacting nuclei distribution which in many cases is a generalized distribution with a proper ionic parameter $q_{\rm i}$ close to one) requires the insertion into the average integral of the penetration factor that is a function of $E$. In the case of the pure Coulomb barrier the screening factor can be factorized. In our general case this factorization is not possible and the behavior of $U_q$ as a function of $D/E$ must be considered with care, both for non-resonant and resonant reactions. The same consideration is valid if instead of $U_q$ we calculate the rates with the use of the plotted and tabulated function $\tau(E)$.

\section{Representative examples of penetration probability}

We report some representative examples on the evaluation of electronic screening factor for non resonant and resonant fusion reactions of interest in solar core and in other dense astrophysical plasmas. Resonant fusions, beside electron screening, are influenced by a resonance screening factor \citep{Cussons}. Furthermore, Maxwellian rates can be corrected by non standard ionic distributions. In this work, we are interested in modified DH potential and here we limit the discussion to electron screening factor.\\
The enhancement of the penetration factor $\Gamma_q(E)$ over the pure Coulomb penetration $\Gamma_{_{\rm C}}(E)$ can be expressed, using Eqs. (\ref{gc}) and (\ref{gq}), by the ratio
\begin{eqnarray}
\nonumber
&&f_{q,\,_{\rm C}}(E)={\Gamma_q(E)\over\Gamma_{_{\rm C}}(E)}\\
\nonumber
&&=\exp\left[-2\,\pi\,{Z_1\,Z_2\,e^2\over\hbar\,c}\,\sqrt{{1\over2}\,\mu\,c^2}\left(
{1\over\sqrt{E+U_q}}-{1\over\sqrt{E}}\right)\right] \ ,\\
&&
\end{eqnarray}
and the enhancement of $\Gamma_q(E)$ over $\Gamma_{_{\rm DH}}(E)\equiv\Gamma_{q\to1}(E)$ by
\begin{eqnarray}
\nonumber
f_{q,\,_{\rm DH}}(E)&=&{\Gamma_q(E)\over\Gamma_{_{\rm DH}}(E)}\\
&=&\exp\left[-2\,\pi\,\eta(E)\,\Big(\tau_q(E)-\tau_{q\to1}(E)\Big)\right] \ .
\end{eqnarray}
It is evident from Figure 2 and from Table 1 that important enhancement over the DH potential comes from the energy range $E\simeq2\cdot10^{-4}\,D\,\div\,D$, with a maximum at about $E\simeq D/15$. The most effective burning energy is at $E_0=(E_{\rm G}\,(k\,T)^2/4)^{2/3}$, where $E_{\rm G}$ is the Gamow energy.\\
\begin{table*}
\caption{Penetration factors for $^{12}$C - $^{12}$C in laboratory experiments with graphite target.}
\vspace{5mm}\begin{center}
\begin{tabular}{lllll}\tableline\tableline
$E\,(MeV)$ \ \ \ & $\Gamma_{_{\rm C}}(E)$ \ \ \ \ \ \ \ \ \ &$\Gamma_{_{\rm DH}}(E)$ \ \ \ \ \ \ \ \ &$\Gamma_{q=0.50}(E)$ \ \ \ \ \ &$\Gamma_{q=0.99}(E)$ \\
\tableline
0.0059 & $8.16\cdot10^{-494}$ & $2.16\cdot10^{-349}$ & $5.20\cdot10^{-285}$ & $1.75\cdot10^{-303}$\\
0.0590 & $1.18\cdot10^{-156}$ & $2.13\cdot10^{-149}$ & $9.70\cdot10^{-141}$ & $2.16\cdot10^{-141}$\\
0.5900 & $4.91\cdot10^{-50}$ & $8.63\cdot10^{-50}$ & $2.00\cdot10^{-49}$ & $1.97\cdot10^{-49}$\\
2.2000 & $2.92\cdot10^{-26}$ & $3.16\cdot10^{-26}$ & $3.56\cdot10^{-26}$ & $3.16\cdot10^{-26}$\\
\tableline
\end{tabular}\end{center}
\end{table*}
The first example concerns the p $-$ p fusion in solar core. At the mass center energies above $E_0=5.9\cdot10^{-3}\,MeV$, $U_q$ can be taken constant and equal to $D=0.7\cdot10^{-4}\,MeV$. The factor $f_{q,\,_{\rm C}}(E)$ is of the order of few percent above unity at high energies, for ions with energy above $E_0$ and belonging to the distribution tail and having fusion probability greater than those belonging to the head of distribution. However, at these high energies $f_{q,\,_{\rm DH}}(E)$ is practically equal to $1$ also with $q\ll1$. Therefore, the rate is not modified significantly respect to the standard value. At energies below $E_0$ the enhancement is not negligible and $U_q$, which depends on $E$, goes to $(1-q)\,D$ as $E\to0$. We have, for instance, $f_{q,\,_{\rm DH}}(E=D)=1.12$ and $f_{q,\,_{\rm DH}}(E=0.1\,D)=1.65$, when $q=0.5$. However, protons with these energies have a very small probability to fuse and the rate in conclusion cannot change respect to the standard evaluation more than few percent, as also required by luminosity constraints.\\
The above discussion is valid for all the other reactions of the hydrogen burning cycle, the effective burning energy being in the range $E_0\simeq15\div30\,KeV$ with $D<E_0$.\\
Among the reactions of CNO cycle we consider $^{14}$N (p, $\gamma$) $^{15}$O. The astrophysical factor recently measured \citep{Runkle,Imbriani} has important consequences in the evolution of stars, estimation of age of globular clusters and of course in the evaluation of CNO neutrino flux \citep{Liolios,Innocenti}. At solar core conditions, we have that in the high energy region $U_q=D\ll E_0=27.0\,KeV$, therefore $\Gamma_q(E)\approx\Gamma_{_{\rm DH}}(E)$. At low energy $U_q\to(1-q)\,D$ and for $E=D/15=0.04\,MeV$ we obtain $\Gamma_q(E)=2.63\cdot10^{-4}$ with $f_{q,\,_{\rm DH}}(E)=3.33$ at $E=0.047\,MeV$ and $q=0.5$. Of course, $f_{q,\,_{\rm DH}}(E)\to1$ in the $q\to1$ limit as we can expect in laboratory experiments.\\
Fusion reactions, like $^7$Li (p, $\alpha$) $\alpha$ and $^6$Li (d, $\alpha$) $\alpha$  are of particular interest for their implications in astrophysics. Experimental measurements in laboratory of the screening potential \citep{Engstler, Pizzone} have indicated a value of $350\div400\,eV$, greater than the adiabatic theoretical one of $186\,eV$. We should evaluate $D$ at the experimental conditions of target temperature and density. The quantity $U_q$ is a fraction of $D$, decreasing to $(1-q)\,D$ and not to zero as in the standard DH approach. Assuming $q=0.5$ we need $D=800\,eV$ to obtain $U_q\simeq400\,eV$. Unfortunately, we do not know the correct target temperature and density to calculate $D$. This difficulty is a motivation to consider questionable the application of this approach to laboratory experiments.\\
We consider now the important non resonant reaction $^{12}$C ($\alpha,\,\gamma$) $^{16}$O at a temperature of $k\,T=17.2\,KeV$ and density $\rho=10^{3.5}\,gr/cm^3$ (as in the core plasma of helium burning red giant stars) with a mass fraction $X($He$)=1/2$ and $X($C$)=1/2$. We have $D=0.573\,KeV$. With $q=0.5$, $U_q$ goes from about $U_q=0.286\,KeV$ at very low energy to $U_q=0.496\,KeV$ at $E=D$ and to $U_q\simeq D$ at high energy. The greatest value of $f_{q,\,_{\rm DH}}(E)$ is at $E\approx4\cdot10^{-2}\,KeV$. At such low energy the screening factor $f_{q,\,_{\rm DH}}(E)\approx10^{55}$, the Coulomb penetration factor being practically zero and $\Gamma_{_{\rm DH}}(E)\approx10^{-458}$.\\
Next we consider the example of $^{12}$C $-$ $^{12}$C fusion in laboratory experiments and in massive stars (in the classical thermonuclear regime, without considering, for simplicity's sake, effects due to the presence of resonances and degeneration \citep{Cussons,Itoh,Ferro1}).
In the experimental study of $^{12}$C $-$ $^{12}$C fusion near the Gamow energy, the target temperature is $k\,T\simeq6\cdot10^{-8}\,MeV$ and the graphite density is $\rho\sim1.7\,gr/cm^3$ \citep{Spillane}, therefore we have approximately $R_{_{\rm DH}}\simeq0.31\cdot10^4\,fm$ and $D\simeq0.0168\,MeV$. From \citep{Assenbaum} we have $U_{\rm e}=5900\,eV$, $R_{\rm a}=0.88\cdot10^4\,fm$, $\Gamma_{_{\rm C}}(E)=\exp(-87.21/\sqrt{E})$ and $\Gamma_{_{\rm DH}}(E)=\exp(-87.21/\sqrt{E+U_{\rm e}})$. We assume $q=0.5$ representing an average deformation of electron distribution. While $U_{\rm e}$ is fixed, $U_q$ depends on energy $E$. Ratios $\Gamma_{_{\rm DH}}(E)/\Gamma_{_{\rm C}}(E)$ coincide with those of Ref. \citep{Spillane}. In Table 2 we report the values of $\Gamma_{_{\rm C}}(E)$, $\Gamma_{_{\rm DH}}(E)$, $\Gamma_{q=0.50}(E)$ and $\Gamma_{q=0.99}(E)$. Our value $\Gamma_{q\to1}(E)$ differs from $\Gamma_{_{\rm DH}}(E)$ because the first factor is calculated using $U_{q\to1}(E)$ and the second using $U_{\rm e}$ fixed.\\
In massive stars \citep{Gasques}, the strong screening effect in a dense plasma can be simulated by assuming $q=0.25$, a value that represents a large electron deformation. At $k\,T=30\cdot10^{-3}\,MeV$ and $\rho\simeq10^9\,gr/cm^3$, we have $R_{_{\rm DH}}\simeq0.5\cdot10^2\,fm$ and $D\simeq1.04\,MeV$. Enhancement of $\Gamma_{q=0.25}(E)$ over $\Gamma_{q=0.99}(E)$ is very evident; at $E=2.2\,MeV$ enhancement is given by a factor of about 2.5. In Table 3 we report, at several energies, $\Gamma_{q=0.25}(E)$ and $\Gamma_{q=0.99}(E)$.\\

\begin{table}
\caption{Penetration factors for
$^{12}$C - $^{12}$C fusion in massive stars.}
\vspace{5mm}\begin{center}
\begin{tabular}{lll}\tableline\tableline
$E\,(MeV)$ \ \ \ \ \ \ &$\Gamma_{q=0.25}(E)$ \ \ \ \ \ \ &$\Gamma_{q=0.99}(E)$\\
\tableline
0.0059 & $1.89\cdot10^{-43}$ & $3.07\cdot10^{-82}$\\
0.0590 & $7.50\cdot10^{-41}$ & $2.42\cdot10^{-54}$\\
0.5900 & $2.19\cdot10^{-31}$ & $1.08\cdot10^{-33}$\\
2.2000 & $6.64\cdot10^{-22}$ & $2.62\cdot10^{-22}$\\
\tableline
\end{tabular}\end{center}
\end{table}

Finally, in Ia supernova environmental conditions give $k\,T=5\cdot10^{-3}\,MeV$, central density $\rho=3\cdot10^9\,gr/cm^3$ and $X($C$)=1/3$. Therefore, $D=1.5\,MeV$, the greatest deviation is at $E=0.1\,MeV$ and consequently we have at this energy $\Gamma_{_{\rm C}}=7.9\cdot10^{-125}$, $\Gamma_{_{\rm DH}}=6.5\cdot10^{-78}$ and $\Gamma_{q=0.5}=4.5\cdot10^{-73}$ with $f_{q,\,_{\rm DH}}=6.9\cdot10^4$.\\
Of course, in all above examples $q$ is an arbitrary parameter whose value should be determined a priori. Rates of fusion reactions will be evaluated using the parametric expression of $U_q(E)$ given in Eq. (\ref{interpo}).

\section{Conclusions}

We have shown that in those systems where in the space coordinates stationary electron distributions deviate from the standard exponential one, the shielded electrostatic Coulomb potential is modified with respect to the standard DH potential derived using linear conditions and constraints. We used two different approaches that produced the same results. One consists of associating a Poisson equation having an unknown electron density distribution with a Bernoulli equation for quantity $r\,V_q(r)$ whose solution is asymptotically a power law.\\
This result can also be obtained, and justified, by means of the super-statistics approach recently developed within generalized statistical mechanics by considering $1/R_{_{\rm DH}}$, the inverse DH radius, as a fluctuating intensive parameter with relative variance $\omega$ and parameter $q$ given by relation $q=1+\omega$, that characterizes the fluctuation.\\
The $q$-modified DH potential has asymptotic power law behavior and we have discussed its meaning for the case $q<1$ which holds when the electrostatic potential has a depleted and cut tail. Fluctuation of the inverse DH radius may be due to temperature and density fluctuation of the electrons surrounding the reacting nuclei in astrophysical plasmas.\\
Exact evaluation of the penetration factor has to be carried out numerically. However, in the case of small deformations, one could calculate $\Gamma(E)$ as the pure Coulomb penetration factor $\Gamma_{_{\rm C}}(E)$ times a correction that may sensibly differ from the standard DH correction.\\ We have reported detailed plots and numerical tables of several useful quantities that allow evaluation of the penetration factor and the reaction rates which, being weighted integrals, depend sensibly on the behavior of the penetration factor as a function of $E$.\\
Penetration factor $\Gamma_q(E)$ can be given as the pure Coulomb penetration factor at an equivalent energy $E+U_q$. Energy $U_q$ is not a constant and can be evaluated only numerically. We have given a useful fit of it. $U_q$ is a function of energy $D$, characteristic of reacting nuclei, their density and fusion temperature, the constant of proportionality depending on $q$ and varying with $D/E$. $U_q$ is proportional to $1/\sqrt{k\,T}$ and to $\sqrt{n}$ as observed experimentally in fusion reactions in metal matrices. The complete and correct expression of the penetration factor or of $U_q$ or $1-\tau_q$ reported in the paper are necessary to evaluate the fusion rates at any energy. In fact, a possible deformed ion distribution together with an electron deformed distribution may sensibly affect the values of the rates. In the evaluations of the rates of fusion reactions, such as in the examples of p $-$ p,  $^{14}$N (p, $\gamma$) $^{15}$O, $^7$Li (p, $\alpha$) $\alpha$ and $^6$Li (d, $\alpha$) $\alpha$, $^{12}$C ($\alpha,\,\gamma)$ $^{16}$O and $^{12}$C $-$ $^{12}$C fusion, mentioned in the previous paragraph, important enhancements may occur, in addition to possible enhancements or decrements due to non-Maxwellian energy-momentum ion distributions, if the electron clouds surrounding the reacting nuclei are spatially modified by an assumed $q<1$ distribution.
Specific applications of our approach are in progress; they are comprehensive of  discussion of the recent experimental results of the LUNA collaboration \citep{Spillane,Imbriani,Gyurky} to derive astrophysical factors at stellar energy range and will be the argument of a further work. We are confident that our approach and numerical tables shown here will be used to study astrophysical process where temperature and density fluctuations cannot be neglected to evaluate DH shielding effect on reaction rates.


\vfill\eject

\begin{thebibliography}{}

\bibitem[Assenbaum et al.(1987)]{Assenbaum} Assenbaum, H. J., Langanke, K.,
\&  Rolfs, C. 1987, Z. Phys. A, 327, 461

\bibitem[Bahcall et al.(1998)]{Bahcall} Bahcall, J., Chen, X., \& Kamionkowski, M. 1998,
\prc, 57, 2756

\bibitem[Beck(2001)]{Beck1} Beck, C. 2001, \prl, 87, 180601

\bibitem[Beck(2004)]{Beck2} Beck, C. 2004, Continuum Mech. Thermodyn., 16, 293

\bibitem[Bellan(2006)]{Bellan}Bellan, P. M. 2006 Foundations of Plasma Physics
(Cambridge University Press, Cambridge)

\bibitem[Bracci et al.(1990)]{Bracci} Bracci, L., Fiorentini, G., Melezhik, V. S.,
 Mezzorani, G., \& Quarati, P. 1990, \nphysa, 513, 316

\bibitem[Bryant(1996)]{Bryant} Bryant, D. A. 1996, J. Plasma Phys., 56, 87

\bibitem[Burlaga et al.(2006)]{Burlaga} Burlaga, L.F., Vi\~{n}as, A.F., Ness, N.F.,
\& Acu\~{n}a, M.H. 2006, \apj, 644, L83

\bibitem[Carraro et al.(1988)]{Carraro} Carraro, C., Schafer, A.,
\& Koonin, S. E. 1988, \apj, 331, 565

\bibitem[Castellani et al.(1997)]{Castellani} Castellani, V., Degl'Innocenti, S.,
Fiorentini, G., Lissia, M., \& Ricci, B., 1997, \physrep, 281, 309

\bibitem[Chevallier(2006)]{Chevallier} Chevallier, L. 2006, Thermalisation of electrons
in a stellar atmosphere, arXiv:astro-ph/0601459v2

\bibitem[Chitanvis(2007)]{Chitanvis} Chitanvis, S. M., \apj, 654, 693

\bibitem[Collins(1989)]{Collins} Collins II, G.W. 1989, The fundamentals of stellar
astrophysics (Freeman, New York)

\bibitem[Coraddu et al.(2004a)]{Coraddu1} Coraddu, M., Lissia, M., Mezzorani, G.,
Petrushevich, Yu. V., Quarati, P., \& Starostin, A. N. 2004, Physica A, 340, 490

\bibitem[Coraddu et al.(2004b)]{Coraddu2} Coraddu, M., Mezzorani, G.,
Petrushevich, Yu. V., Quarati, P., \& Starostin, A. N. 2004, Physica A, 340, 496

\bibitem[Coraddu et al.(2006)]{Coraddu3} Coraddu, M., Lissia, M., Mezzorani, G.,
\& Quarati, P. 2006, Eur. J. Phys. B, 50, 11

\bibitem[Cravens(1997)]{Cravens} Cravens, T. E. 1997, Physics of Solar System Plasmas
(Cambridge University Press, Cambridge)

\bibitem[Cussons et al.(2002)]{Cussons} Cussons, R., Langanke, K., \& Liolios, T. 2002,
Eur. Phys. J. A, 15, 291

\bibitem[Degl'Innocenti et al.(2004)]{Innocenti} Degl'Innocenti, S., et al. 2004,
Phys. Lett. B, 590, 13

\bibitem[Engstler et al.(1992)]{Engstler} Engstler, S., et al. 1992, Zeit. Phys. A, 342, 471

\bibitem[Ferro(2004)]{Ferro1} Ferro, F., Lavagno, A., \& Quarati, P. 2004,
Eur. Phys. J. A, 21, 529

\bibitem[Ferro \& Quarati(2005)]{Ferro} Ferro, F., \& Quarati, P. 2005, \pre, 71, 026408

\bibitem[Gasques et al.(2005)]{Gasques} Gasques, L. R., et al. 2005, \prc, 72, 025806

\bibitem[Gruzinov \& Bahcall(1998)]{Gruzinov} Gruzinov, A. V., \& Bahcall, J. 1998,
\apj, 504, 996

\bibitem[Gyurky(2007)]{Gyurky} Gyurky, Gy., et al. 2007, $^3$He ($\alpha,\,\gamma$)
$^7$Be cross section at low energies, arXiv:nucl-ex/0702003

\bibitem[Ichimaru(1993)]{Ichimaru} Ichimaru, S. 1993, Rev. Mod. Phys., 65, 255

\bibitem[Imbriani et al.(2005)]{Imbriani} Imbriani, G., et al. 2005, Eur. Phys. J. A,
25, 455

\bibitem[Itoh et al.(2003)]{Itoh} Itoh, N., Tomizawa, N., Wanajo, S., \& Nozawa, S.
2003, \apj, 586, 1436

\bibitem[Johnson et al.(1992)]{Johnson} Johnson, C. W., Kolbe, E., Koonin, S. E.,
\& Langanke, K. 1992 \apj, 392, 320

\bibitem[Kim \& Jung(2004)]{Kim1} Kim, C. -G., \& Jung, Y. -D. 2004,
Plasma Phys. Contr. Fusion, 46, 1493

\bibitem[Kim \& Zubarev(2006)]{Kim} Kim, Y. E., \& Zubarev, A. L. 2006,
Japanese J. Appl. Phys., 45, L552

\bibitem[Leubner(2004)]{Leubner} Leubner, M. P. 2004, \apj, 604, 469

\bibitem[Leubner(2005)]{Leubner2} Leubner, M. P. 2005, \apj, 632, L1

\bibitem[Leubner \& V\"or\"os(2005)]{Leubner1} Leubner, M.P., \& V\"or\"os, Z. 2005,
\apj, 618, 547

\bibitem[Liolios(2000)]{Liolios} Liolios, T. E.  2000, \prc, 61, 055802

\bibitem[Lissia \& Quarati(2005)]{Lissia} Lissia, M., \& Quarati, P. 2005,
Europhysics News, 36, 211

\bibitem[Maero et al.(2006)]{Maero} Maero, G., Quarati, P., \& Ferro, F. 2006,
Eur. Phys. J. B, 50, 23

\bibitem[Opher \& Opher(2000)]{Opher} Opher, M., \& Opher, R. 2000, \apj, 535, 473

\bibitem[Oxenius(1986)]{Oxenius} Oxenius, J. 1986, Kinetic Theory of Particles
and Photons, Theoretical foundations of non LTE Plasma Spectroscopy (Sprinter Verlag, Berlin)

\bibitem[Peyraud-Cuenca(1992)]{Peyraund} Peyraud-Cuenca, N. 1992, \aap, 261, 633

\bibitem[Pizzone et al.(2003)]{Pizzone} Pizzone, R. G., et al. 2003, \aap, 398, 423

\bibitem[Raiola et al.(2004)]{Raiola} Raiola, F., et al. 2004,
Eur. Phys. J. A, 19, 283

\bibitem[Ricci et al.(1995)]{Ricci} Ricci, B., Degl'Innocenti, S., \& Fiorentini, G.
1995, \prc, 52, 1095

\bibitem[Rolfs \& Rodney(2005)]{Rolfs} Rolfs, C.E., \& Rodney, W.S. 2005,
Cauldrons in the Cosmos: Nuclear Astrophysics (University of Chicago Press)

\bibitem[Rubab \& Murtaza(2006)]{Rubab} Rubab, N., \& Murtaza, G. 2006,
\physscr, 74, 145

\bibitem[Runkle et al.(2005)]{Runkle} Runkle, R. C., et al. 2005, \prl, 94, 082503

\bibitem[Salpeter(1953)]{Salpeter1} Salpeter, E. E. 1953,
Australian J. Phys., 7, 373

\bibitem[Salpeter \& Van Horn(1969)]{Salpeter2} Salpeter, E. E.,
\& Van Horn, H. M. 1969, \apj, 155, 183

\bibitem[Shaviv(2004)]{Shaviv1} Shaviv, G. 2004, Prog. Theor. Phys. Suppl., 154, 293

\bibitem[Shaviv et al.(2000)]{Shaviv2} Shaviv, G., \& Shaviv, N. J. 2000, \apj,
529, 1054

\bibitem[Shaviv et al.(2001)]{Shaviv3} Shaviv, G., \& Shaviv, N. J. 2001, \apj,
558, 925

\bibitem[Shaviv et al.(2002)]{Shaviv4} Shaviv, G., \& Shaviv, N. J. 2002, AIP
Conf. Proceed., 637, 150

\bibitem[Shoppa et al.(1993)]{Shoppa} Shoppa, T.D., Koonin, S.E., Langanke, K.,
\& Seki, R. 1993, \prc, 48, 837

\bibitem[Spillane et al.(2007)]{Spillane} Spillane, T.,  et al. 2007,
Study of the $^{12}$C - $^{12}$C fusion reactions near the Gamow energy, arXiv:nucl-ex/0702023

\bibitem[Strieder et al.(2001)]{Strieder} Strieder, F., Rolfs, C., Spitaleri, C.,
\& Corvisiero, P. 2001, Naturwissenschaften, 88, 461

\bibitem[Treuman et al.(2004)]{Treuman} Treuman, R. A., Jaroschek, C. H.,
\& Scholer, M. 2004, Phys. Plasmas, 11, 1317

\bibitem[Tsallis(1988)]{Tsallis1} Tsallis, C. 1988, J. Stat. Phys., 52, 479

\bibitem[Tsallis \& Borges(2003)]{Tsallis2} Tsallis, C., \& Borges, E.P. 2003,
Nonextensive statistical mechanics - Applications to nuclear and high energy physics, Proceedings of the Xth International Workshop on Multiparticle Production - Correlations and Fluctuations in QCD, Antoniou  N., ed. (World Scientific, Singapore); arXiv:cond-mat/0301521

\bibitem[Wilk \& W{\l}odarczyk(2001)]{Wilk2} Wilk, G., \& W{\l}odarczyk, Z. 2001,
Chaos, Solitons and Fractals, 13, 581

\bibitem[Wilk(2006)]{Wilk1} Wilk, G., \& W{\l}odarczyk, Z.  2006, Fluctuations, correlations and
non-extensivity, arXiv:hep-ph/0610292v3

\end{thebibliography}
\end{document}